\documentclass{article}

\usepackage{url}
\usepackage{listings}

\usepackage{common}

\begin{document}

\title{\normalfont\bfseries\large Robust Topological Orderings for Directed Graphs}
\author{James Smith\\\texttt{james.smith@djalbat.com}}
\date{}
	
\maketitle

\begin{abstract}
\noindent We modify the Pearce-Kelly algorithm that maintains a topological ordering for a directed acyclic graph in order to allow cycles to be tolerated. 
Cycles make topological orderings moot, of course, however tolerating them is useful in practice.
A user may mistakenly introduce a cyclic dependency in their project,, for example, and then subsequently fix their mistake.
In these cases it is better to maintain the relevant data structures so that if and when the directed graph becomes acyclic again, a topological ordering can be instantly recovered.
It turns out that adding this functionality costs us little, only small modifications and some attention to detail are needed.
\end{abstract}

\section{Introduction}

Algorithms for dealing with directed graphs, and indeed many other kinds of algorithms that deal with data, fall broadly into two categories.  Batch algorithms take an initial and complete input of data and create the necessary data structures for output. They will probably process their input and return their output efficiently,, however if the data changes then they need to be run again in their entirety. This is likely to be far from optimal in practice. Incremental algorithms, on the other hand, store their data structures in such as way as to allow them to be maintained in the face of any subsequent partial data inputs.

The Pearce-Kelly algorithm~\cite{Pearce-Kelly} is an incremental algorithm for maintaining topological orderings of directed acyclic graphs. The way it works is explained in the next section, and for now we simply point out one way in which it can be modified, namely making it robust in the face of cycles. In this way if subsequent inputs result in these cycles being removed then the algorithm will once again output a topological ordering. We outline these modifications in what follows, and since they really require no more than an understanding of how the Pearce-Kelly algorithm works, we review that in detail first.
\section{How the Pearce-Kelly algorithm works}

\begin{figure}[H]
\centering
\includegraphics[scale=0.75]{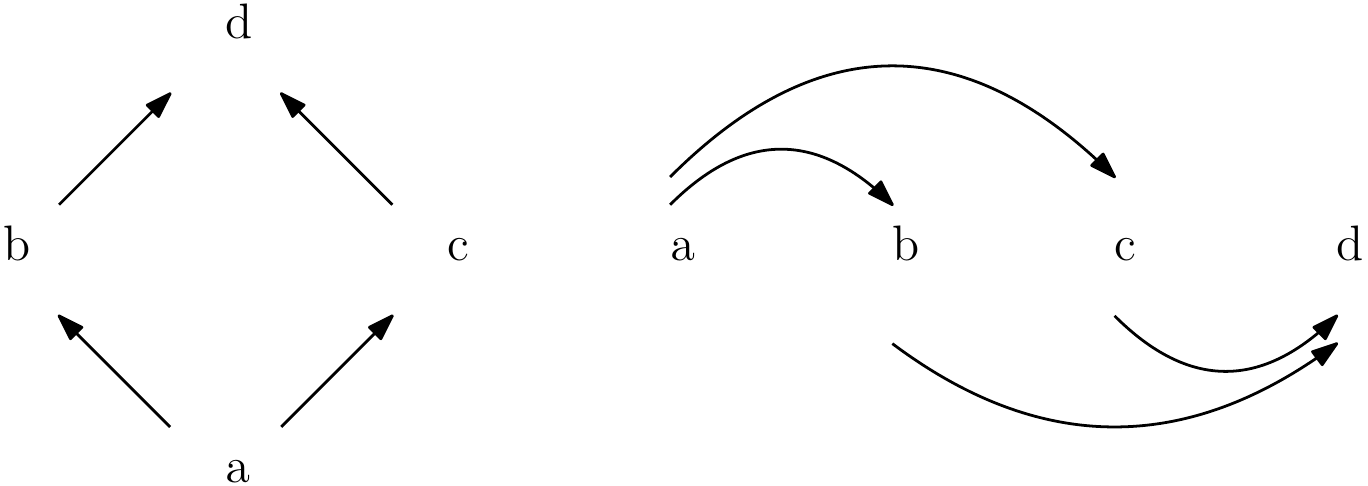} 
\caption{An acyclic directed graph versus one of its topological orderings}
\label{graph-versus-ordering}
\end{figure}

Consider a directed acyclic graph and one of its topological orderings shown in figure~\ref{graph-versus-ordering}.

\noindent Conceptually we tend to view such graphs in two dimensions, as on the left,  but in fact the ordering on the right, if complete with arrows as shown, represents the same graph. The main insight behind the Pearce-Kelly algorithm is to store the graph in this form and, as a consequence, in order to return a topological ordering it has to do no more than return its own data structure minus the arrows. The only requirement aside from this is to work out whether or not an edge can be added without creating a cycle and, if it can be added, rearrange the vertices in such as way as to maintain a topological ordering whilst accommodating the new edge.

\begin{figure}[H]
\centering
\includegraphics[scale=0.75]{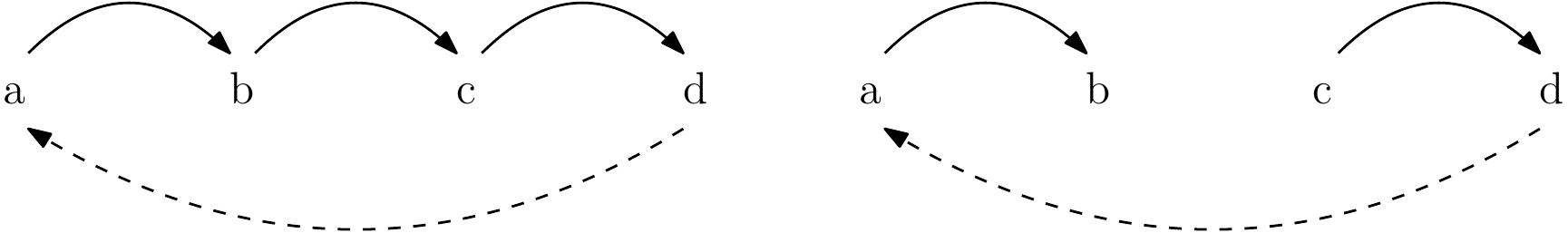} 
\caption{Cyclic versus non-cyclic edges about to be added}
\label{cyclic-versus-non-cyclic-edge}
\end{figure}

\noindent In figure~\ref{cyclic-versus-non-cyclic-edge} two directed graphs are shown with an edge being added to each. On the left, adding the edge would result in a cycle whereas on the right it would not. What is a suitable test to make this intuition precise? In fact all we need to do is to check whether or not the source vertex of the edge to be added is reachable from the target vertex. If it is then adding the edge would create a cycle.  To check for reachability a standard depth first search is employed.

\begin{figure}[H]
\centering
\includegraphics[scale=0.75]{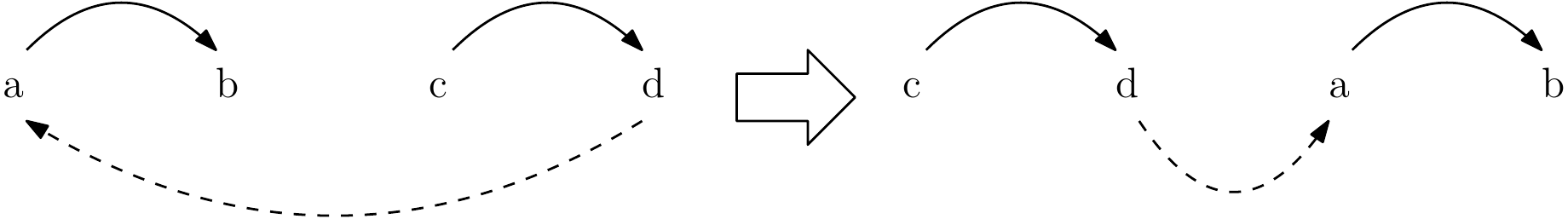} 
\caption{A simple reordering of vertices in order to accommodate an additional edge}
\label{simple-reordering}
\end{figure}

On the right we may proceed with adding the edge. As already mentioned, a feature of the Pearce-Kelly algorithm is that it stores the graph as a topological ordering and what this means is that all the arrows go from left to right. Therefore when we rearrange the vertices in order to accommodate the new edge this property of the arrows must persist. The process is shown in figure~\ref{simple-reordering}. To explain it briefly, all of the vertices reachable from the target vertex going forwards together with the target vertex itself are swapped with all of the vertices reachable from the source vertex going backwards together with the source vertex. The result is that the additional edge now also goes from left to right.

\begin{figure}[H]
\centering
\includegraphics[scale=0.75]{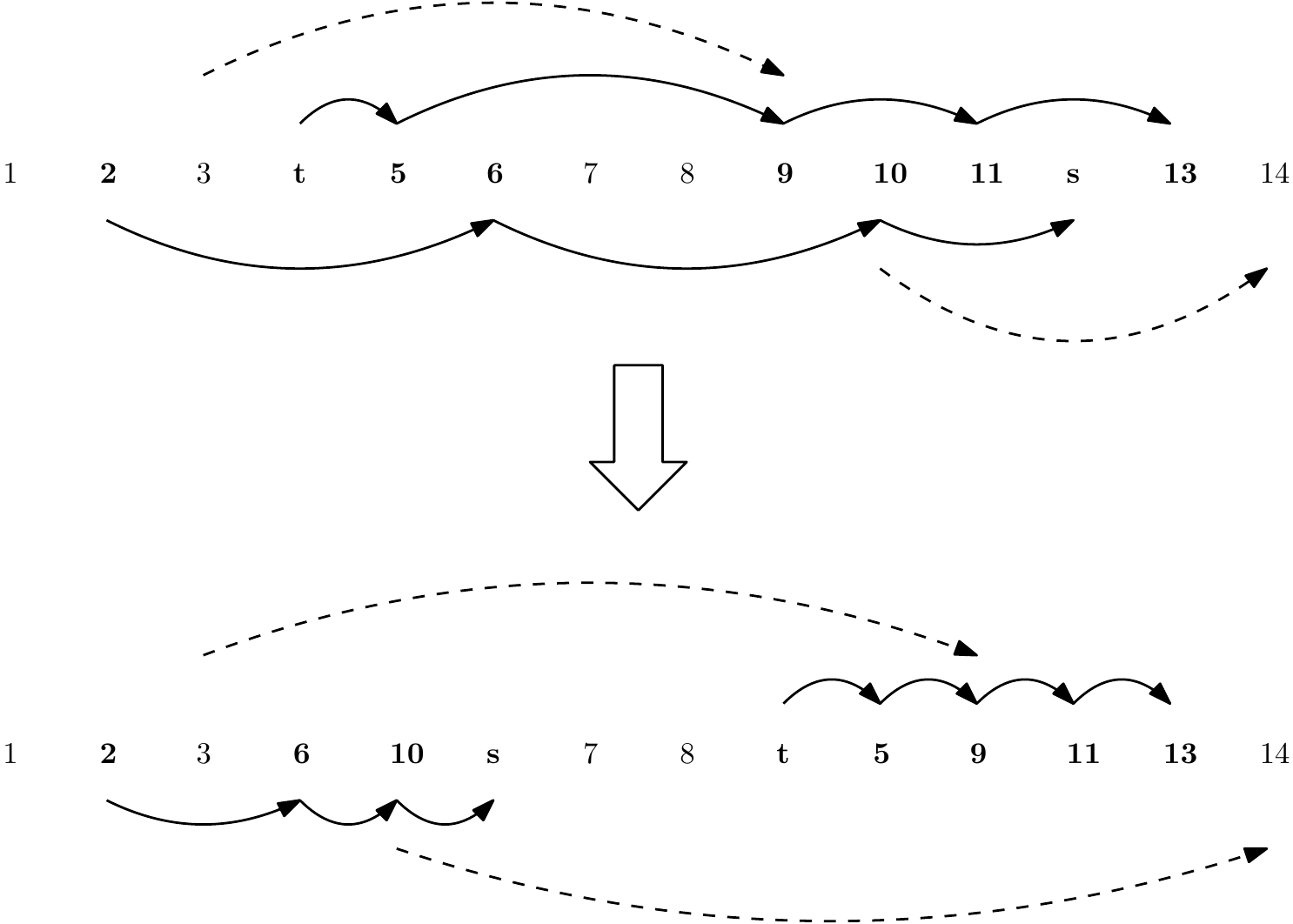} 
\caption{A more complex reordering of vertices in order to accommodate an additional edge}
\label{complex-reordering}
\end{figure}

\noindent In fact a more complex case is worth treating because precisely why the algorithm works in all cases has a considerable bearing on the modifications that we will make to it. Figure~\ref{complex-reordering} shows such a case and we make the following points about it:\\

\begin{itemize}
\item Consider the vertices reachable from either of the additional edge's vertices, namely the vertices \{ 2, t, 5, 6, 9, 10, s, 13 \}. These are interspersed amongst other vertices, as would likely be the case in practice.  
\item Consider the vertices reachable from the target vertex of the additional edge going forwards together with the target vertex itself,, namely the vertices \{ t, 5, 9, 11, 13 \}. Let us call them the target vertices. Note that there are no other vertices that are reachable from these vertices going forwards. If there were, they would also be one of these vertices by definition.
\item Similarly, consider the vertices reachable from the source vertex of the additional edge going backwards together with the source vertex itself,, namely the vertices \{ 2, 6, 10, s \}. Let us call them the source vertices. Note that there are no other vertices that are reachable from these vertices going backwards. If there were, they would also be one of these vertices by definition.
\item The collections of source vertices and target vertices are distinct. If they were not then there would be a path from the target vertex to the source vertex and therefore the additional edge would create a cycle.\\
\end{itemize}

\noindent Now consider any edge into the target vertices. From the points above and from figure~\ref{complex-reordering} it should be clear that it will only ever be stretched to the right by the reordering and can never be reversed. Similarly any edge out of the source vertices will only ever be stretched by the reordering and can never be reversed.

\noindent It only remains to mention that when an edge is removed from the graph no reordering is necessary because the topological ordering will obviously be preserved. Next we describe our modifications to the algorithm,. bearing all of the above in mind.

\section{Modifications to the algorithm}

There are several points that need addressing when adding support for cyclic edges. Firstly,. how should we define a cyclic edge? Secondly, how does the presence of cyclic edges affect reachability? Thirdly, what happens when we add an edge to a directed graph that has cycles? And lastly,  what happens when we remove an edge?

\begin{figure}[H]
\centering
\includegraphics[scale=0.75]{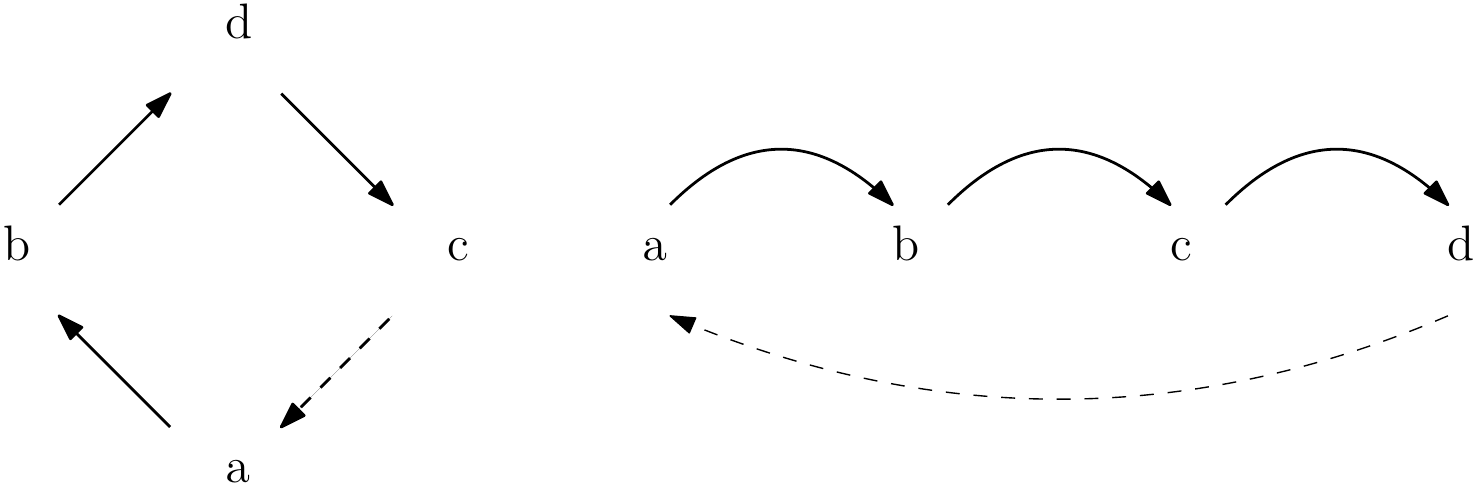} 
\caption{A directed graph versus one of its topological representations}
\label{graph-versus-representation}
\end{figure}

In order to address the first point we note that if we are to support cycles then we must add the edge on the left irregardless of the fact that a cycle results. Therefore we simply add it as-is, with its target vertex coming before its source vertex. And this constitutes our definition of a cyclic edge, in fact, namely one that goes from right to left. Figure~\ref{graph-versus-representation} gives an example. We cannot consider the data structure on the right as being akin to a topological ordering any more and therefore we refer to it as a topological representation. It should be clear that if any topological representation has an arrow going from right to left then the corresponding directed graph has a cycle and vice versa.

\noindent Any topological representation can conceptually be split into two. On the one hand we have the edges that go from left to right that contribute to the underlying directly acyclic graph and on the other hand we have what we term the cyclic edges that always go from right to left and result in cycles in the overarching directed graph. Of course any one of the edges in a cycle can be considered as cyclic but we stick with defining only those edges that go from right to left in the topological representation as being cyclic. We also stress the observation that we can recover the underlying directly acyclic graph from the overarching directed graph by discounting these cyclic edges since it only takes the removal of one edge of a cycle to remove the cycle itself. 

Coming to the second point, we modify our depth first search to only take into account non-cyclic edges, that is edges that go from left to right in the topological representation. Consequently reachability will only pertain to the underlying directly acyclic graph. The utility of this approach will become apparent in what follows.

For the third point the question has now become not simply whether on not we can add an edge to a directly acyclic graph without creating a cycle but whether or not we can add an edge to a directed graph that might already have cycles without creating another. We choose to dodge this question, however, in effect only asking the original question. We can do this because our depth first search ignores cyclic edges and therefore effectively only traverses the underlying directly acyclic graph, but we need to justify this decision. We argue as follows. Suppose that there are no cycles. Then there is no difference between the underlying and overarching graphs and the algorithm can proceed as before. Now suppose that cycles do exist. We note that adding another edge, whether itself cyclic or not, has no effect on these existing cycles.We can only affect these cycles by removing edges, never by adding them.  Therefore by attempting to add an edge to the underlying graph we do not affect the overarching graph other than by potentially adding another cycle to it if it is impossible to avoid adding a cycle to the underlying graph. In short,adding an edge to the underlying graph does not effect the outcomes that would arise from trying to add an edge to the overarching graph.

Finally, in order to address the fourth point we modify the algorithm to re-examine each of the cyclic edges whenever an edge is removed. For each cyclic edge the algorithm again checks whether or not the edge's source vertex is reachable from its target vertex in the underlying directed graph. If it is not then the edge is removed and can be added as a non-cyclic edge to the underlying directed acyclic graph in the standard way with its attendant reordering. 

Thus the gist of our approach is to treat the underlying directly acyclic graph wherever possible by modifying out depth first search to only take non-cyclic edges into account. Edges are then added in much the same way, the only difference being that they are added regardless of whether a reordering is possible; and when edges are removed there is a subsequent filtering of the extant cyclic edges.

\section{Conclusions}

We have modified the Pearce-Kelly algorithm in order to make it robust in the face of cycles in the directed graph. The modifications are few. Firstly, we restrict depth first searches to the underlying directed acyclic graph in all cases. Secondly, we allow edges to be added whether they are cyclic or not and lastly, we check extant cyclic edges to see if they have become non-cyclic whenever an edge is removed. This last modification will slow the algorithm down when edges are removed but otherwise the modified algorithm will perform just as well as the original. A JavaScript implementation can found found here~\cite{Occam:Directed-Graphs}.

\pagebreak

\bibliographystyle{plain}
\bibliography{references}

\end{document}